\documentclass[12pt]{article}
\usepackage{fullpage}
\usepackage{amsmath}
\usepackage{amssymb}
\usepackage[]{graphicx}
\graphicspath{{graphics/}} % Sets the default location of pictures

\usepackage{epstopdf}
\usepackage{multirow}
\usepackage{color}
\usepackage[linktocpage]{hyperref}
\usepackage{cite}
\usepackage{wrapfig}
\usepackage{subeqnarray}
\usepackage{lscape}
\usepackage[mathscr]{eucal}
\usepackage{dcolumn}
\newcolumntype{d}[1]{D{.}{\cdot}{#1} }

\usepackage{upgreek}
\newcommand{\SImum}{\ensuremath{\upmu}\textrm{m}}
\def\blue{\textcolor{black}}

\def\##1{{\bf #1}}
\def\=#1{\underline{\underline #1}}

\def\.{\mbox{ \tiny{$^\bullet$} }}

\def\eps{\varepsilon}

\def\epsa{\eps_{a}}
\def\epsb{\eps_{b}}
\def\epsc{\eps_{c}}

\def\ux{\hat{\#u}_x}
\def\uy{\hat{\#u}_y}
\def\uz{\hat{\#u}_z}
\def\uprop{\hat{\#u}_{prop}}

\def\ped0{_{\scriptscriptstyle 0}}
\def\muo{\mu_{\scriptscriptstyle 0}}
\def\epso{\eps_{\scriptscriptstyle 0}}
\def\etao{\eta_{\scriptscriptstyle 0}}
\def\lambdao{\lambda_{\scriptscriptstyle 0}}
\def\ko{k_{\scriptscriptstyle 0}}
\def\co{c_{\scriptscriptstyle 0}}
\def\tq{\tilde{q}}

\def\vph{v_{ph}}
\def\propdist{\Delta_{prop}}

\def\fref#1{Fig.~\ref{#1}}

\def\sref#1{Sec.~\ref{#1}}
\def\tref#1{Table~\ref{#1}}

\def\tond#1{\left(#1\right)}
\def\quadr#1{\left[#1\right]}

\begin{document}

\large
{\bf Temperature-mediated transition from Dyakonov--Tamm surface waves to surface-plasmon-polariton waves}
\normalsize

\vspace{10pt}
 
Francesco Chiadini$^{1}$,
Vincenzo Fiumara$^{2}$,
Tom G. Mackay$^{3,4}$,
Antonio Scaglione$^{1}$, and
Akhlesh Lakhtakia$^{4}$

\vspace{10pt}

$^{1}$Department of Industrial Engineering,
	University of Salerno, via Giovanni Paolo II, 132 -- Fisciano (SA),
	84084, Italy;
	
	 $^{2}$School of Engineering, University of
	Basilicata, Viale
	dell'Ateneo Lucano 10, 85100 Potenza, Italy;
	
$^{3}$School of Mathematics and
   Maxwell Institute for Mathematical Sciences,
University of Edinburgh, Edinburgh EH9 3FD, UK;

 $^{4}$Department of Engineering Science and Mechanics, Pennsylvania State University,
	University Park, PA 16802--6812,
	USA
	
%\ead{T.Mackay@ed.ac.uk}

\vspace{10pt}

\begin{abstract}
The effect of changing the temperature on the propagation of electromagnetic surface waves (ESWs), guided by the planar interface of a  homogeneous isotropic temperature-sensitive  material (namely, InSb) and a  temperature-insensitive structurally chiral material (SCM)  was numerically investigated  in the terahertz frequency regime. As the temperature rises,  InSb  transforms  from  a dissipative dielectric material to a \blue{dissipative}  plasmonic material. 
Correspondingly,  the ESWs transmute from Dyakonov--Tamm surface waves into surface--plasmon--polariton  waves. The effects of the temperature change are clearly observed in the 
 phase speeds, propagation distances, angular existence domains,  multiplicity, and spatial profiles of energy flow  of the ESWs. Remarkably  large propagation distances can be achieved;
   in such instances
 the  energy of an ESW is confined almost entirely  within the SCM. For certain propagation directions,   simultaneous excitation of  two ESWs with (i) the same phase speeds but different propagation distances or (ii) 
 the same propagation distances but different phase speeds
  are also indicated by our results.
\end{abstract}
\vspace{2pc}
\vspace{2pc}

%\submitto{\JOPT}

%\maketitle

%\ioptwocol

\section{Introduction} \label{sec:intro}

The propagation of any electromagnetic surface wave (ESW) is guided by the planar interface  of two different mediums
\cite{Boardman,PMLbook}.  
Uller proved theoretically in 1903 that the two different mediums can be isotropic and homogeneous dielectric materials,
the relative permittivities of both partnering materials being positive real, so long as at least one of the two is dissipative.
The  prediction was theoretically confirmed shortly thereafter, first
by Zenneck \cite{Zenneck} and then by Sommerfeld \cite{Sommerfeld1,Sommerfeld2,Sommerfeld3}, and experimentally during this decade \cite{FLol}.

A major development came in 1977, when
ESWs  were theoretically predicted   to exist at
the interface of two isotropic dielectric materials \cite{YYH},
at least one of which is periodically nonhomogeneous in
the direction normal to the interface. These ESWs were
named Tamm waves as they are analogous to the electronic
states predicted to exist at the surface of a crystal
by Tamm in 1932 \cite{Tamm}. Tamm waves have been experimentally
observed \cite{YYC}, and even applied for sensing
purposes \cite{SR2005,KA2007,Sinibaldi,KKAVSD}.

If the periodically nonhomogeneous partnering material
is anisotropic, the ESWs are called Dyakonov--Tamm (DT)  surface waves \cite{LP2007}.
The periodic nonhomogeneity may be either piecewise homogeneous or continuous.
As an example, a Reusch pile \cite{Reusch,Joly1,Joly2,Joly3,Joly4} supports DT surface-wave propagation \cite{DT_Reusch_pile}. This material is piecewise homogeneous  as it is a stack of anisotropic dielectric layers with an incremental rotation from one
layer to the next about an axis normal to the layers. An equichiral Reusch pile has only two layers per period, whereas an ambichiral Reusch pile has are more than two layers per period \cite{ambi}. If the number of layers per period is sufficiently large, the Reusch pile is classified as finely chiral. A very finely chiral Reusch pile  may be regarded as a
structurally chiral material (SCM)~\cite{Bose,Chandra,DeG,AkhMes} whose nonhomogeneity is effectively continuous.

The allowed directions of propagation in the interface plane define the angular existence domain (AED)
of ESWs. The AED  of DT surface waves is  often so large as
to  encompass the entire interface plane. In contrast, if the anisotropic and periodically
nonhomogeneous partnering material were to be made homogeneous, the AED shrinks to a few
degrees in width, if that \cite{Takayama1,Takayama2}. Hence, DT surface waves are expected to
be exploited for several applications including optical sensing \cite{LFjnp}.
Furthermore, the multiplicity of DT surface waves that can be excited at a specific frequency for a given propagation direction~---~as predicted by theory \cite{LP2007} and later observed in experimental studies \cite{PML2013,PMLHL2014}~---~is very appealing because it can enhance the sensitivity as well as the reliability of sensing and also allow for the simultaneous  detection of multiple analytes.

 The temperature dependence of the relative permittivity $\eps_{th}$
of certain isotropic dielectric materials, e.g., InSb in the terahertz frequency regime, allows them to change from slightly dissipative dielectric materials (i.e., $\mbox{Re}(\eps_{th}) > 0$) to plasmonic materials
(i.e., $\mbox{Re}(\eps_{th}) < 0$) or \textit{vice versa}, with the transition occurring at  a 
temperature for which $\mbox{Re}(\eps_{th}) = 0$~\cite{Howells}.
If such a homogeneous material were to partner a
Reusch pile or a SCM, then DT surface waves could be  transformed into
surface-plasmon-polariton (SPP) waves \cite{PMLbook,Pitarke,Maier},  or \textit{vice versa}. 

SPP waves are guided by the planar
interface of a plasmonic material (often, a metal at optical frequencies) and a dielectric material that
can be either isotropic \cite{Pitarke} or anisotropic \cite{Sprokel} and either homogeneous \cite{Pitarke}
or periodically nonhomogeneous \cite{PMLbook}. Due to their widespread exploitation for optical sensing
\cite{Homola,SL2016}, SPP waves have an extensive literature. SPP waves also have applications for
near-field microscopy \cite{Kawata,BindingEvents}, harvesting solar energy \cite{Mokka,Anderson}, and 
communications \cite{Sekhon}. In AED and multiplicity, SPP waves and DT surface waves have similar characteristics
\cite{PMLbook,ChJOSAB}.

In this paper, we focus on the  transmutation of DT surface waves into  SPP waves when a critical change
in temperature transmutes the homogeneous partnering material (InSb) from a dissipative
dielectric material  to a \blue{dissipative} plasmonic material, the nonhomogeneous  anisotropic
partnering material being an SCM. In \sref{sec:matmeth},  we briefly present the theoretical preliminaries for the canonical boundary-value problem that describes ESWs guided by the planar interface of the two chosen materials.
Comprehensive details of the solution procedure are presented elsewhere \cite{LP2007,PL2009,ChOC,ChJNP}.
 In \sref{sec:nrd}
we report on the characteristics of the excited ESWs  in terms of their wavenumbers, phase speeds, propagation distances,  AEDs, and  spatial profiles of the 
 time-averaged Poynting vector. By comparing the characteristics of the  ESWs when $\mbox{Re}(\eps_{th})$ for InSb is 
 positive and negative,  we characterize the features of the transmutation  from DT surface waves to SPP waves. Conclusions follow in \sref{sec:cr}.

An $\exp\tond{-i\omega t}$  dependence on time $t$ is implicit, with $\omega$ denoting the angular frequency and $i=\sqrt{-1}$. The free-space wavenumber, the free-space wavelength, and the intrinsic impedance of free space are denoted by $\ko=\omega\sqrt{\epso \muo}$, $\lambdao=2\pi/\ko$, and $\etao=\sqrt{\muo/\epso}$, respectively, with $\epso$ and $\muo$ being the permeability and permittivity of free space. The speed of light in
free space is denoted by $\co=1/\sqrt{\epso\muo}$. Vectors are in boldface; dyadics are underlined twice;   and Cartesian unit vectors are identified as $\ux$, $\uy$, and $\uz$.

\section{Theoretical Preliminaries}\label{sec:matmeth}
In order to  investigate temperature-sensitive ESW propagation guided by the planar interface of InSb (or a similar homogeneous
material)
and a SCM, we formulated a canonical boundary-value problem,
and thereby obtained a dispersion equation that  was numerically solved.

\begin{figure}[h!]
	\centering
	\includegraphics[width=0.8\linewidth]{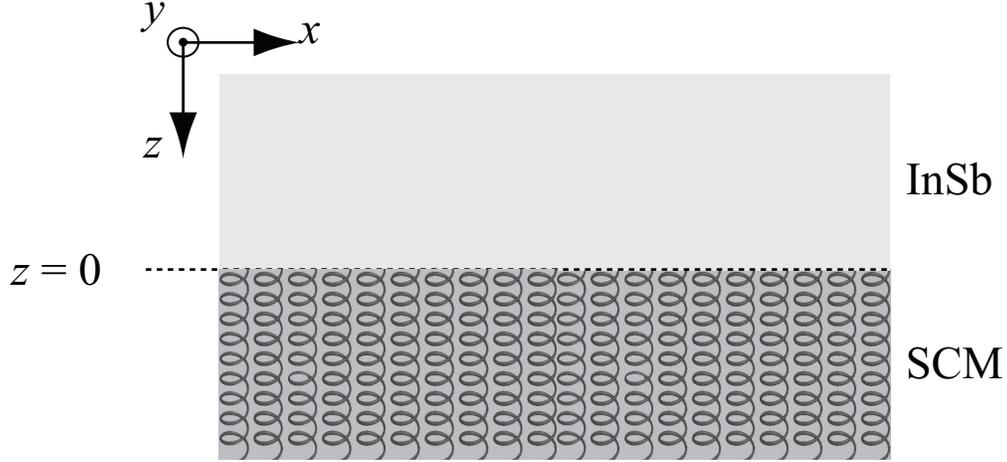}
	\caption{Schematic of the  canonical boundary-value problem solved.}
	\label{Fig1}
\end{figure}

A schematic of the canonical  boundary-value problem  is provided in  Fig.~\ref{Fig1}. The half~space $z<0$ is occupied by a homogeneous and isotropic material,   namely InSb, whose relative permittivity $\eps_{th}$ varies with temperature in a known way.
\blue{The half~space $z>0$
is occupied by a SCM which is assumed to be a unidirectionally nonhomogeneous material, characterized by constitutive parameters that vary continuously and periodically along the $z$ direction \cite{AkhMes}. This macroscopic assumption is valid provided that 
the length scale of the SCM's morphology is considerably smaller than the 
electromagnetic wavelengths involved. By a process of local homogenization \cite{AkhMes}, the macroscopic description can be inferred from the
  underlying nanostructure of the  SCM \cite{ML_Sensors_J}.
The
 nonhomogeneous relative permittivity dyadic of the SCM is given by}
\begin{equation}
\=\eps_{\rm SCM}\left(z\right)= \=S_z(z)\cdot\=S_{y}(\chi) \cdot \=\eps^{\circ} _{ref}\cdot \=S_{y}^{-1}(\chi) \cdot \=S_{z}^{-1}(z)\,,
\label{perm_SCM}
\end{equation}
with the  dyadic
\begin{equation}
\=\eps^{\circ} _{ref} =\epsa \,\uz\uz+\epsb \,\ux\ux+\epsc \,\uy\uy\,
\label{epsref}
\end{equation}
indicating local orthorhombicity.
 Herein  $\epsa$, $\epsb$, and $\epsc$ are complex-valued, $\omega$-dependent, parameters whose values are assumed to be
independent of temperature. The $\omega$-independent rotation and tilt dyadics
\begin{equation}
\left.\begin{array}{l}
\=S_{z}\left(z\right)=(\ux\ux+\uy\uy) \cos\left(\pi{z}/{\Omega}\right)\\
\quad +h\,
(\uy\ux - \ux\uy) \sin \left(\pi{z}/{\Omega}\right) +\uz\uz
\\[5pt]
\=S_{y}(\chi) =(\ux\ux+\uz\uz) \cos\chi \\
\quad +
(\uz\ux - \ux\uz) \sin \chi +\uy\uy
\end{array}\right\}
\label{perm_SCM_dyadics}
\end{equation}
together
delineate the SCM's helicoidal morphology with $h=1$ denoting structural right-handedness and $h=-1$  structural left-handedness,
 $2\Omega$ is the structural period of the SCM along the $z$ axis, and $\chi\in(0,\pi/2]$.
 Both partnering materials are assumed to have unit relative permeability.

%%%%%%%%%%%%%%%%%% Figure 1 begins %%%%%%%%%%%%%%%%
%%%%%%%%%%%%%%%%%% Figure 1 ends %%%%%%%%%%%%%%%%

 We consider the ESW to be  propagating parallel to the unit vector $\uprop=\ux\cos\psi+\uy\sin\psi$, $\psi\in[0,2\pi)$, in the  $xy$ plane.  The magnitudes of the field phasors of the ESW decay to zero as   $z \to \pm \infty$. With  $q$ as   the   wavenumber, the electric and magnetic  phasors  of the ESW  can be represented everywhere  by
\begin{equation}
\left.\begin{array}{l}
 \#E(\#r)= \#e(z)  \, \exp\left({iq}\uprop\cdot\#r\right)\\
\#H(\#r)= \#h(z) \, \exp\left({iq}\uprop\cdot\#r\right)
\end{array}\right\}
\label{eq:EH1}
\end{equation}
where $\#r$ is the   position vector. Appropriate representations of the amplitude functions $\#e(z)$ and $\#h(z)$ in the half~spaces
$z<0$ and $z>0$, and the derivation of the corresponding dispersion relation and  subsequent  extraction of $q$ therefrom, are comprehensively described elsewhere
\cite{PMLbook,LP2007,PL2009,ChOC,ChJNP}. For each value of $\psi$, the dispersion relation may yield  multiple  values of $q$.

\section{Numerical Results and Discussion}\label{sec:nrd}

We numerically solved the dispersion relation to obtain the normalized wavenumbers $\tq =q/k\ped0 $ of the ESWs.
Once  $q$ (or $\tq$) is known,  the corresponding propagation length and phase speed of the ESW were calculated as
$\propdist = 1 /\mbox{Im} (q)$ and
 $ \vph=\co/ {\rm Re}\tond{\tq}$, respectively.
The  spatial  profile  of the rate of energy flow  associated with an  ESW is provided via the time-averaged Poynting vector $\#P(\#r)=(1/2)\,{\rm Re}\left[\#e(z) \times \#h^\ast(z)\right]\exp[-2{\rm Im}(q)\uprop\cdot\#r]$, where the asterisk denotes the complex conjugate.
For all numerical results reported here, we fixed $\lambda\ped0=500$~\SImum~and  $\Omega=200$~\SImum, while  the direction of propagation was  varied in the
entire $xy$ plane.  

The  partnering material occupying the half-space $z<0$  was taken
to be the semiconductor InSb, whose relative permittivity in the terahertz regime is given by the Drude model \cite{Howells,Han}
\begin{equation}
 \eps_{th} =  \eps_\infty - \frac{\omega_p^2}{\omega^2 + i \gamma \omega },
\end{equation}
wherein the high-frequency relative permittivity $\eps_\infty = 15.68$, the damping constant $\gamma = \pi \times 10^{11}$~rad~s$^{-1}$, and the plasma frequency $\omega_p = \sqrt{N q_e^2 / 0.015\, \epso \, m_e}$ depends upon the electronic charge $q_e =
-1.60 \times 10^{-19}$~C and mass $m_e = 9.11 \times 10^{-31}$~kg. The  dependence of $ \eps_{th} $ on temperature $T$ (in K) is mediated by the intrinsic carrier density (in m$^{-3}$) \cite{Gruber,Zimpel,Halevi}
\begin{equation}
N = 5.76 \times 10^{20} \, T^{3/2} \, \exp \left( - \frac{{\sf E_g}}{2 k_B T} \right)\,,
\end{equation}
with ${\sf E_g}=0.26$~eV being the band-gap energy and
$k_B = 8.62 \times 10^{-5}$ eV $\mbox{K}^{-1}$ being the Boltzmann constant.
Whereas
$\eps_{th} =  10.95 + 0.39 i  $ at   $T= 180$~K, $\eps_{th}=  -13.66 + 2.44 i$ at
$T= 220$~K. Thus,   InSb is a dissipative  dielectric material at $T= 180$~K but a metal at $T= 220$~K,
with ${\rm Re}\left(\eps_{th}\right)=0$ at $T\approx 204.56$~K.
The constitutive parameters of the SCM were chosen to be $\eps_a=1.574$, $\eps_b=3.228$, $\eps_c=2.313$, $\chi=20^\circ$, and $h=1$.

%%%%%%%%%%%%%%%%% Figure 2 begins %%%%%%%%%%%%%%%%
\begin{figure}[h]
	\centering
	\includegraphics[width=0.8\linewidth]{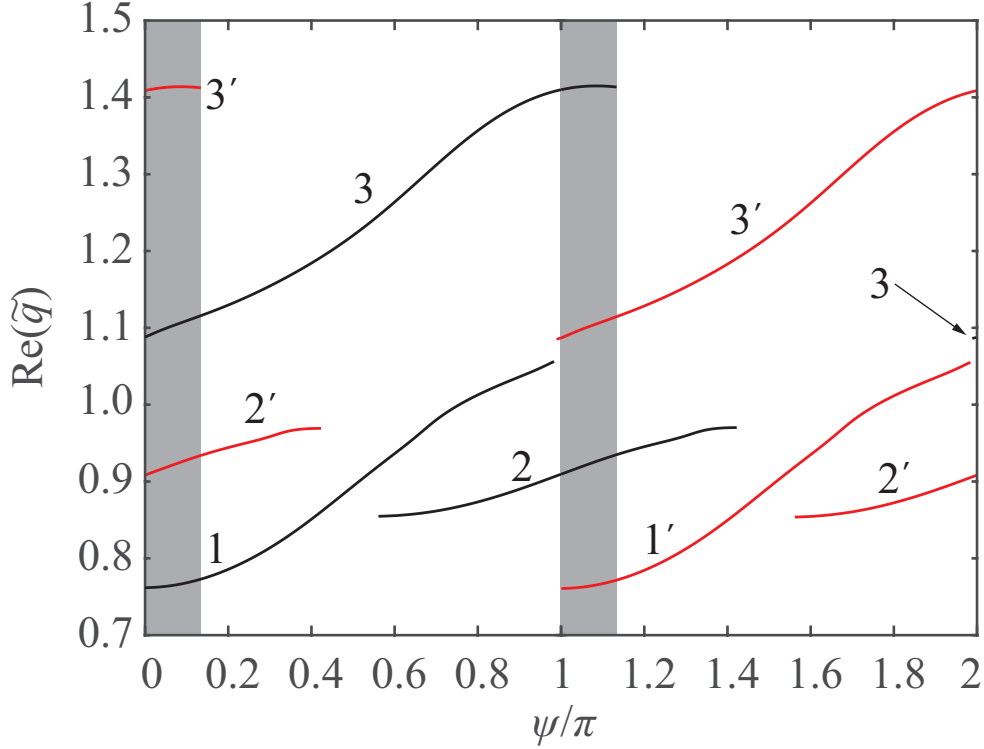}
	\caption{Variation of ${\rm Re}(\tq)$   with $\psi$ of DT surface waves  guided by the planar InSb/SCM interface at $T=180$~K ($\eps_{th} =  10.95 + 0.39i$). The solutions of the dispersion equation are organized in $6$ numbered
	branches. Ranges of $\psi$ with four ESWs are shaded gray in Figs.~\ref{fig:reqT180}--\ref{fig:propdT200},
	\ref{fig:reqT220}, and \ref{fig:propdT220}.}
	\label{fig:reqT180}
\end{figure}
%%%%%%%%%%%%%%%%% Figure 2 ends %%%%%%%%%%%%%%%%

%%%%%%%%%%%%%%%%% Figure 3 begins %%%%%%%%%%%%%%%%
\begin{figure}[h]
	\centering
	\includegraphics[width=0.8\linewidth]{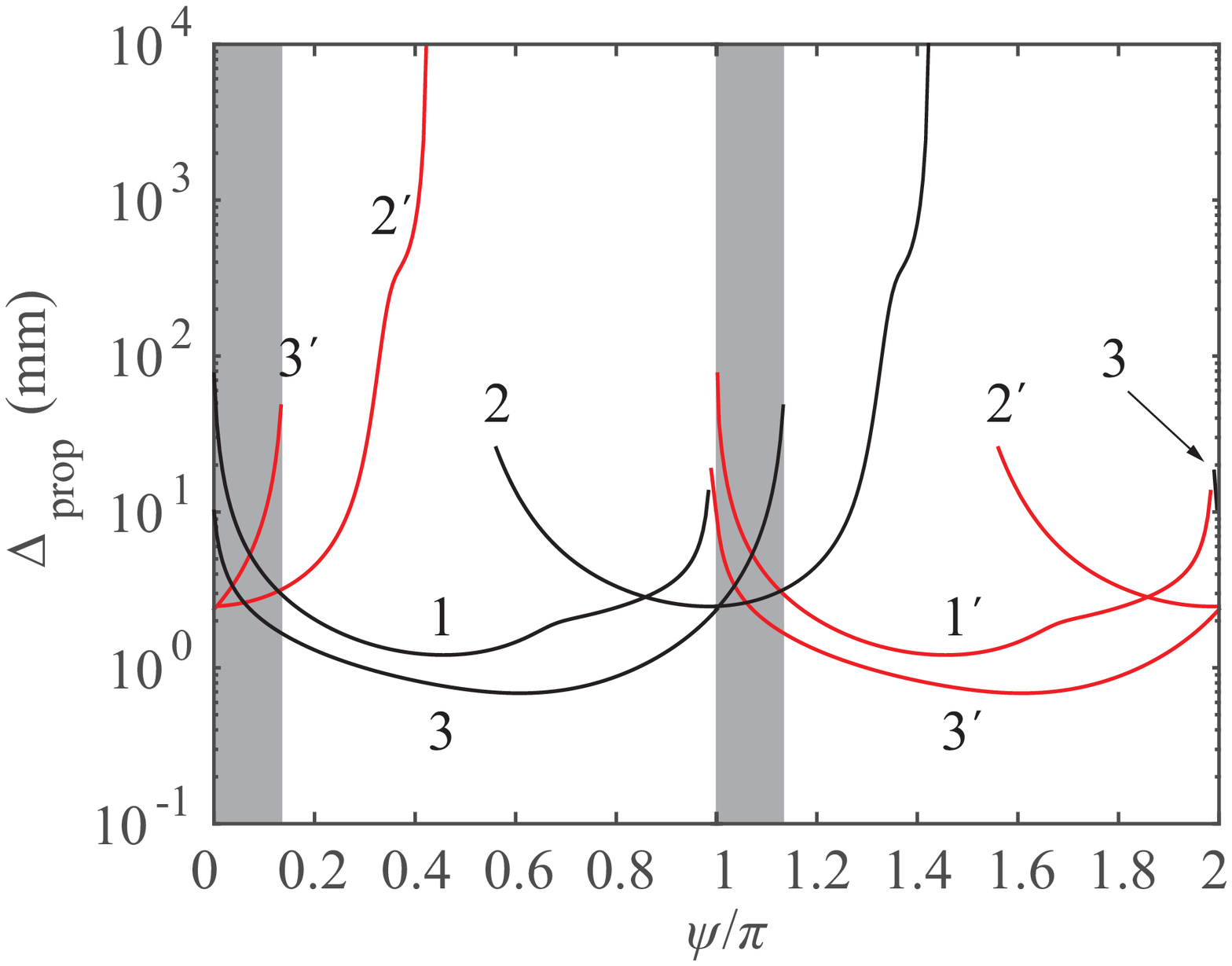}
	\caption{As Fig.~\ref{fig:reqT180} but the variation of $\propdist$ is presented. }
	\label{fig:propdT180}
\end{figure}
%%%%%%%%%%%%%%%%% Figure 3 ends %%%%%%%%%%%%%%%%

\subsection{Wavenumbers and propagation distances}\label{ESWT}
Figures~\ref{fig:reqT180} and \ref{fig:propdT180},  respectively, show the real part of the normalized wavenumber $\tq$ and the propagation distance $\propdist$
for all solutions of the dispersion relation, as the angle $\psi$ of the direction of propagation  in the $xy$ plane is varied from 0 to $ 2 \pi$, with $T=180$~K fixed.  The solutions were organized into  two sets each composed by three different  branches numbered
$1$ to $3$, and $1^\prime$ to $3^\prime$, respectively. If $q$ is a solution for a specific value of the propagation angle $\psi$ on branch $\ell$, it is also a solution for either $\psi+\pi$ or $\psi-\pi$ on branch $\ell^\prime$. This symmetry arises from that  of $\=\eps_{\rm SCM}\left(z\right)$  in the $xy$ plane.

All  branches in Figs.~\ref{fig:reqT180} and \ref{fig:propdT180} span limited ranges of $\psi$  that in some cases overlap, either partially or totally.  These overlaps indicate
 the  multiplicity of DT surface waves  for the ranges of $\psi$ involved. For any propagation angle, the interface guides at least two DT surface waves. Furthermore,
 as many as four DT surface waves can propagate in the intervals $0\leq \psi \leq 0.1333\pi$ and  $1\leq \psi \leq 1.1333\pi$, both of which are   highlighted in gray in  Figs.~\ref{fig:reqT180} and \ref{fig:propdT180}.

As the propagation angle $\psi$ increases, the value of ${\rm Re}(\tq)$ in Figs.~\ref{fig:reqT180} increase on  branches $1$, $2$, $1^\prime$, and $2^\prime$. The same is also true on branch $3$
for $\psi<1.1111\pi$ and on branch $3^\prime$ for  $\psi<0.1111\pi$ and $\psi>0.994\pi$, with  
${\rm Re}(\tq)$ slightly decreasing as $\psi$ increases for $1.1111 \pi <\psi < 1.133 \pi$ on branch $3$ and for
 $0.1111 \pi <\psi < 0.133 \pi$   on branch  $3^\prime$. The phase speeds of the DT surface waves
on branches $2$ and $2^\prime$, on branch $1$ for $0\leq \psi <  0.7556\pi$,  and on branch $1^\prime$ for 
$1\leq \psi <  1.7556\pi$ are higher than the speed of light in vacuum (i.e., $v_{ph}>\co $).  In contrast, the phase speeds of the DT surface waves
on branches $3$ and $3^\prime$, on branch $1$ for $0.7556< \psi \leq  0.9830\pi$,  and on branch $1^\prime$ for $1.7556< \psi \leq  1.9830\pi$ are lower than the speed of light in vacuum (i.e., $v_{ph}<\co $).

At six different values of $\psi$, 
 two branches intersect in \fref{fig:propdT180} but not in
\fref{fig:reqT180}.  Consequently, for each of the six intersections in \fref{fig:propdT180},
 a pair of DT surface waves can be simultaneously excited  with the same attenuation rates in the interface plane but with different phase speeds.

Data on AEDs, relative phase speeds,  and propagation distances  are reported in \tref{tab:T180}. The maximum value of $\propdist$ in this table is a phenomenally large 9.95~m on  branches 2 and $2^\prime$~---~two orders of magnitude greater than the maximum propagation distances on any of the other branches.

%%%%%%%%%%% BEGIN TABLE 1 %%%%%%%%%%%%%%%%%%%%%%%%%%%%%%%%%%
\begin{table*}[h]
	\caption{Angular existence domain, minimum and maximum values of relative phase speed $\vph/\co$, and minimum and maximum values of propagation distance $\propdist$ for DT surface waves guided by the InSb/SCM interface when $T=180$~K.}
	\centering
	\begin{tabular}{c|cccd{2}d{2}}
		\hline\hline
		Branch & AED ($\psi/\pi$) & $\vph^{\tond{min}}/\co$ & $\vph^{\tond{max}}/\co$ &
		  \multicolumn{1}{c}{$\propdist^{\tond{min}}$~(mm)} &  \multicolumn{1}{c}{$\propdist^{\tond{max}}$~(mm)}\\ [0.5ex]
		\hline
		$ 1\ $ & $\quadr{0,0.983}$ & $0.9468$ & $1.3127$ & $1.21$ &$76.52$ \\ \hline
		$ 1^\prime$ & $\quadr{1,1.983}$ & $0.9468$ & $1.3127$ & $1.21$ &$76.52$ \\ \hline
		$ 2\ $ & $\quadr{0.561,1.422}$ & $1.0308$ & $1.1700$ & $2.48$ &$9947.18$ \\ \hline
		$ 2^\prime$ & $\quadr{0,0.422}\cup\quadr{1.561,2}$ & $1.0308$ & $1.1700$ & $2.48$ &$9947.18$ \\ \hline
		$ 3\ $ & $\quadr{0,1.133}\cup\quadr{1.994,2}$ & $0.7093$ & $0.9198$ & $0.69$&$19.40$ \\ \hline
		$ 3^\prime$ & $\quadr{0,0.133}\cup\quadr{0.994,2}$     & $0.7093$ & $0.9198$ & $0.69$ &$19.40$ \\ \hline
	\end{tabular}
	\label{tab:T180}
\end{table*}
%%%%%%%%%%% END TABLE 1 %%%%%%%%%%%%%%%%%%%%%%%%%%%%%%%%%%%%

%%%%%%%%%%%%%%%%% Figure 4 begins %%%%%%%%%%%%%%%%
\begin{figure}[h]
	\centering
	\includegraphics[width=0.8\linewidth]{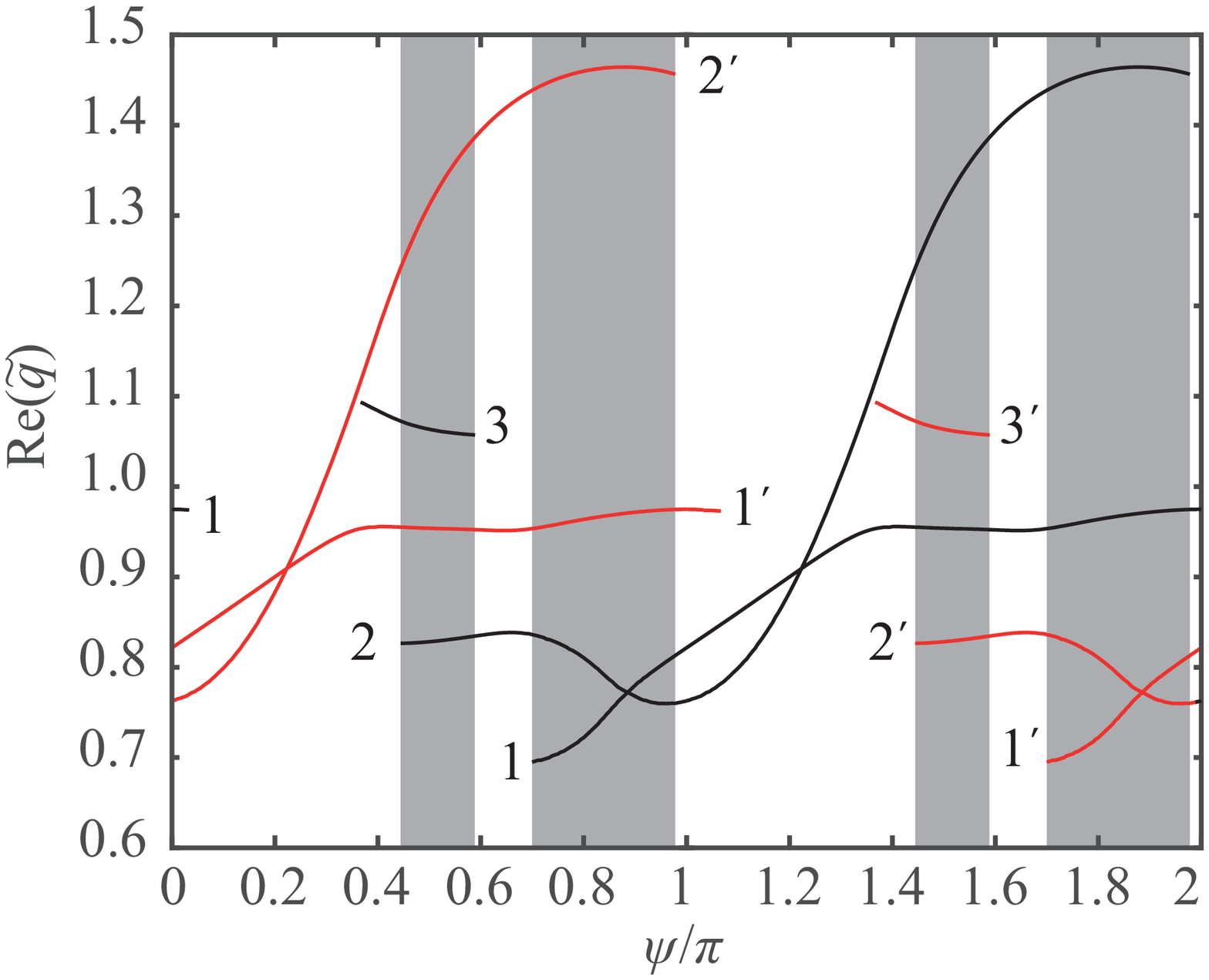}
	\caption{As \fref{fig:reqT180} except that  $T=200$~K ($\eps_{th} =  2.87 + 1.07
		i  $).}
	\label{fig:reqT200}
\end{figure}
%%%%%%%%%%%%%%%%% Figure 4 ends %%%%%%%%%%%%%%%%

%%%%%%%%%%%%%%%%% Figure 5 begins %%%%%%%%%%%%%%%%
\begin{figure}[h]
	\centering
	\includegraphics[width=0.8\linewidth]{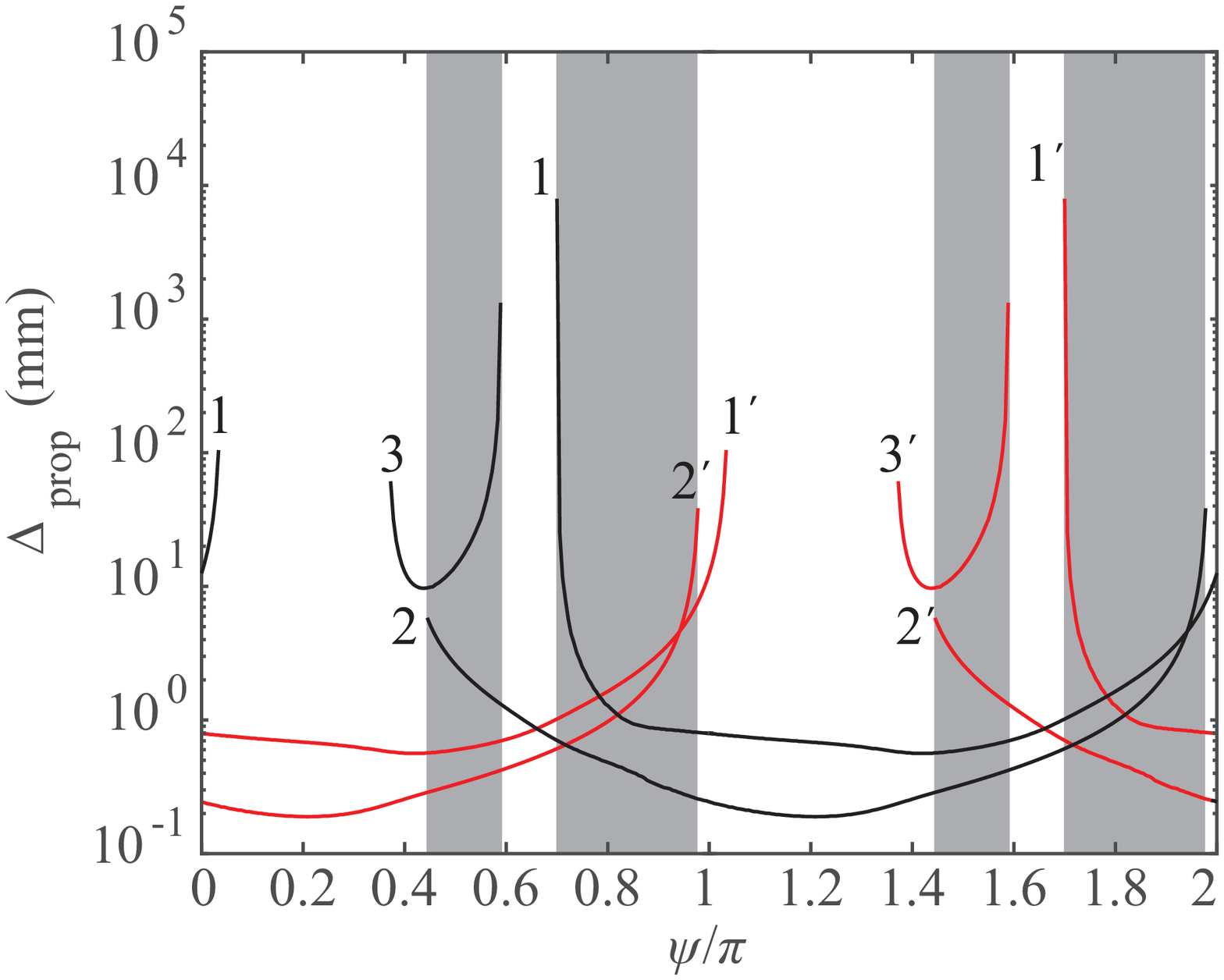}
	\caption{ As \fref{fig:propdT180} except that $T=200$~K ($\eps_{th} =  2.87 + 1.07
		i  $).}
	\label{fig:propdT200}
\end{figure}
%%%%%%%%%%%%%%%%% Figure 5 ends %%%%%%%%%%%%%%%%

The real part of the normalized
wavenumber and the propagation distance, respectively,   are presented in Figs.~\ref{fig:reqT200} and 
\ref{fig:propdT200} as $\psi/\pi$ varies  for  ESWs  guided by  the planar InSb/SCM interface when  $T=200$~K. These ESWs are classified as DT surface waves because $\eps_{th} =  2.87 + 1.07i  $.  A  comparison  with  Figs.~\ref{fig:reqT180} and \ref{fig:propdT180}  reveals that increasing the temperature
 by $20$~K  does not affect the number of  solution branches, there being six  solution branches  at  $200$~K.  The  branches are numbered from
$1$ to $3$ and  from $1^\prime$ to $3^\prime$.  If $q$ is a solution for a specific value of $\psi$ on branch $\ell$, it is also a solution for either $\psi+\pi$ or $\psi-\pi$ on branch $\ell^\prime$.

All six branches at $T=200$~K cover limited  ranges of $\psi$.  Except for branches 3 and $3^\prime$ for which ${\rm Re}(\tq)$ decreases as the propagation angle $\psi$ increases, no monotonic trend can be recognized in the other branches.
On  branches $1$ and $1^\prime$, on branch $2$ for $0.444\pi\leq \psi <  1.289 \pi$, and on branch $2^\prime$ for $0\leq \psi <  0.289 \pi$ and $1.444\leq \psi \leq  2 \pi$, the  phase speeds exceed  $\co$; in contrast,  $v_{ph}<\co $ on branch $2$ for $1.289< \psi\leq 1.978\pi$, and  on branch $2^\prime$ for $0.289< \psi\leq 0.978\pi$, as well as on branches $3$ and $3^\prime$.

The overlaps of different branches results in a multiplicity of DT surface waves that can be as high as four, depending upon $\psi$. The four ranges where  the maximum multiplicity of DT surface waves  arises ($0.444\pi\leq\psi\leq 0.589\pi $, $0.7\pi \leq\psi \leq 0.978\pi$, $1.444\pi\leq\psi\leq 1.589 \pi$, and $1.7\pi \leq\psi \leq 1.978 \pi $) are highlighted in gray in Figs.~\ref{fig:reqT200} and \ref{fig:propdT200}. 

Figure~\ref{fig:reqT200} reveals an interesting phenomenon. Branches $1^\prime$ and $2^\prime$ intersect at $\psi =0.2232\pi$ and $\psi =1.8855\pi$, whereas branches $1$ and $2$ intersect at $\psi =0.8855\pi$ and $\psi =1.2232\pi$. None of the four intersections is found in \fref{fig:propdT200}. Hence,
at each of the four values of $\psi$ identified, two DT surface waves can be simultaneously excited  with the same phase speeds but  different attenuation rates in the interface plane. 

Two branches intersect at ten different values of $\psi$ in
\fref{fig:propdT200}. None of those ten intersections are evident in
\fref{fig:reqT200}. Hence, at each of ten intersections in \fref{fig:propdT200},
 a pair of DT surface waves can be simultaneously excited  with the same attenuation rates in the interface plane but with different phase speeds.

Data on AEDs, relative phase speeds,  and propagation distances at $T=200$~K are provided in \tref{tab:T200}.
The maximum propagation distance on branches 1  and $1^\prime$ is a remarkably large $7.96$~m.

%%%%%%%%%%% BEGIN TABLE 2 %%%%%%%%%%%%%%%%%%%%%%%%%%%%%%%%%%
\begin{table*}[h]
	\caption{Angular existence domain, minimum and maximum values  of normalized phase speed $\vph/\co$, and minimum and maximum values of propagation distance $\propdist$ for DT surface waves guided by the InSb/SCM interface when $T=200$~K.}
	\centering
	\begin{tabular}{c|cccd{2}d{2}}
		\hline\hline
		Branch & AED ($\psi/\pi$) & $\vph^{\tond{min}}/\co$ & $\vph^{\tond{max}}/\co$ &
		  \multicolumn{1}{c}{$\propdist^{\tond{min}}$~(mm)} &  \multicolumn{1}{c}{$\propdist^{\tond{max}}$~(mm)}\\ [0.5ex]
		\hline
		$ 1$ & $\quadr{0,0.033}\cup\quadr{0.700,2}$ 	 & $1.0259$ & $1.4388$ & $0.56$  &$7957.75$ \\ \hline
		$ 1^\prime$ & $\quadr{0,1.033}\cup\quadr{1.700,2}$ 	 & $1.0259$ & $1.4388$ & $0.56$  &$7957.75$ \\ \hline
		$ 2$ & $\quadr{0.444,1.978}$ 	 & $0.6830$ & $1.3162$ & $0.19$  &$38.44$ \\ \hline
		$ 2^\prime$ & $\quadr{0,0.978}\cup\quadr{1.444,2}$ 	 & $0.6830$ & $1.3162$ & $0.19$  &$38.44$ \\ \hline
		$ 3$ & $\quadr{0.372,0.589}$ & $0.9147$ & $0.9461$ & $9.69$  &$1326.29$ \\ \hline
		$ 3^\prime$ & $\quadr{1.372,1.589}$ & $0.9147$ & $0.9461$ & $9.69$  &$1326.29$ \\ \hline
	\end{tabular}
	\label{tab:T200}
\end{table*}
%%%%%%%%%%% END TABLE 2 %%%%%%%%%%%%%%%%%%%%%%%%%%%%%%%%%%%%

%%%%%%%%%%%%%%%%% Figure 6 begins %%%%%%%%%%%%%%%%
\begin{figure}[h!]
	\centering
	\includegraphics[width=0.8\linewidth]{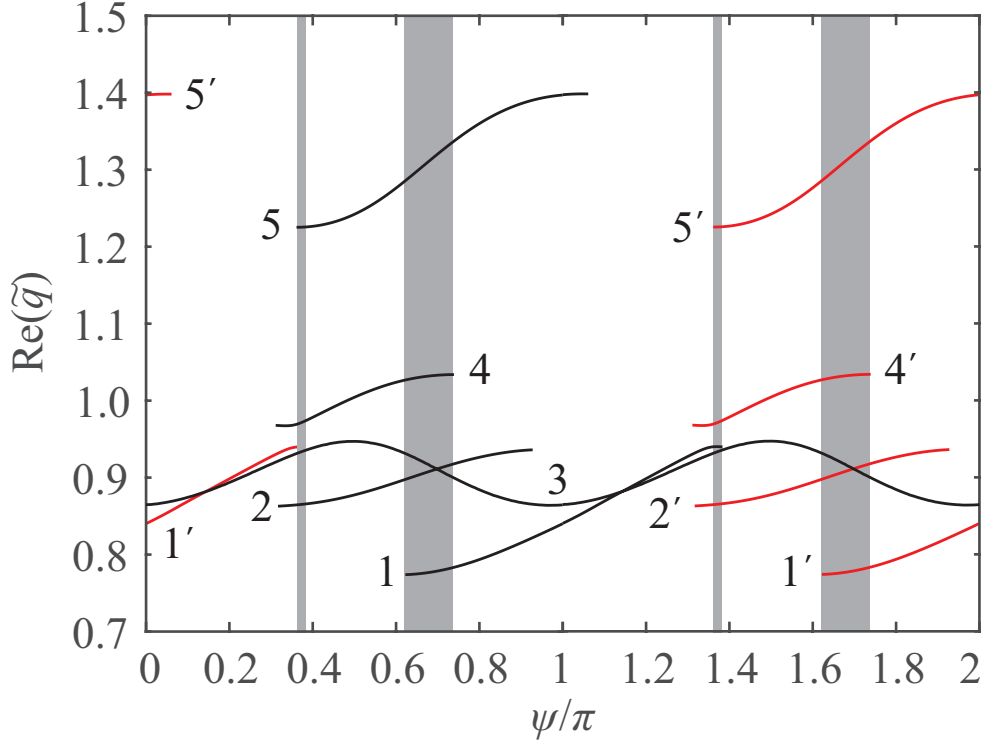}
	\caption{
As \fref{fig:reqT180} except that  $T=205$~K ($\eps_{th} =   -0.30+1.33
		i  $). Ranges of $\psi$ with 5  SPP  waves are shaded gray.
}
	\label{fig:reqT205}
\end{figure}
%%%%%%%%%%%%%%%%% Figure 6 ends %%%%%%%%%%%%%%%%

%%%%%%%%%%%%%%%%% Figure 7 begins %%%%%%%%%%%%%%%%
\begin{figure}[h!]
	\centering
	\includegraphics[width=0.8\linewidth]{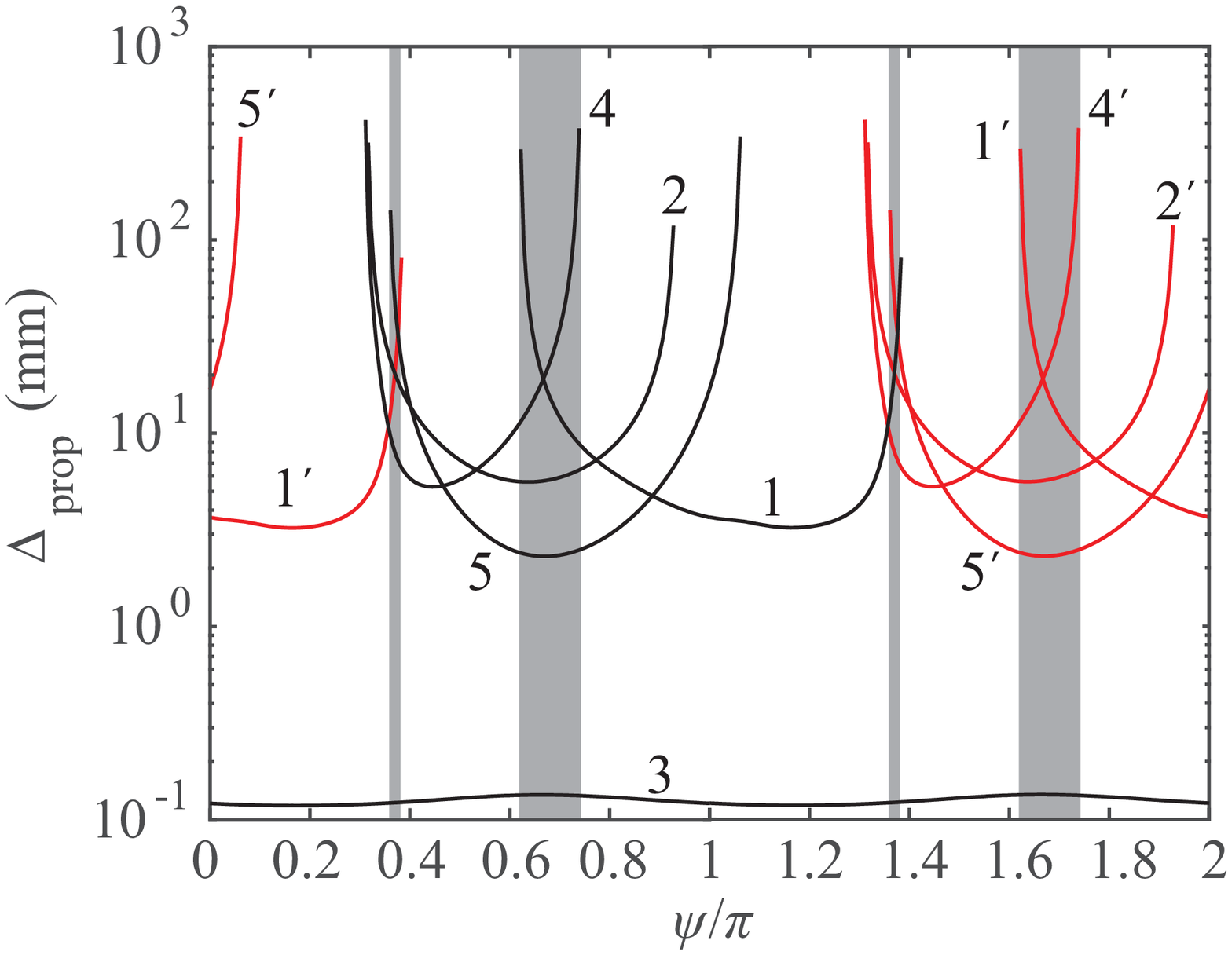}
	\caption{
As \fref{fig:reqT205}   but the variation of $\propdist$ is presented.
}
	\label{fig:propdT205}
\end{figure}
%%%%%%%%%%%%%%%%% Figure 7 ends %%%%%%%%%%%%%%%%

Figures~\ref{fig:reqT205} and \ref{fig:propdT205} show the variations of
 ${\rm Re}(\tq)$ and $\propdist$, respectively,   with
$\psi$   when  $T=205$~K.
The increment of $5$~K from 200~K
 results in the sign of  $\mbox{Re} (\eps_{th})$ changing from positive to negative, with  ${\rm Re}(\eps_{th})=0$ \blue{occurring} at $T\approx 204.56$~K. Since InSb changes from being  a dissipative dielectric material to a \blue{dissipative} plasmonic material,
 a comparison of Figs.~\ref{fig:reqT200} and \ref{fig:propdT200} on the one hand with Figs.~\ref{fig:reqT205} and \ref{fig:propdT205} on the other hand
reveals the effects  of the transition from DT surface waves to SPP waves.
The number of solution branches increases from six (numbered $1$ to $3$ and $1^\prime$ to $3^\prime$) at 200~K to nine (numbered $1$ to $5$, $1^\prime$, $2^\prime$, $4^\prime$, and $5^\prime$) at  $205$~K. If $q$ is a solution for a particular value of $\psi$, it is also a solution for $\psi\pm\pi$.

The propensity for multiple solutions of the dispersion equation enhances with the temperature rising to 205~K, with
as many as five SPP waves existing in four angular ranges: $0.361\pi\leq \psi \leq 0.383\pi$, $0.622\pi\leq \psi \leq 0.739\pi$, 
$1.361\pi\leq \psi \leq 1.383\pi$, and $1.622\pi\leq\psi\leq1.739\pi$. In contrast to the case for $T=200$~K and $T=180$~K, when $T=205$~K one branch (namely, branch $3$) spans the entire angular range $0\leq \psi<2\pi$. 

Two SPP waves can be simultaneously excited at 205~K with the same phase speed but not identical propagation distances at: (\textit{i}) $\psi=0.143\pi$ where branches  $1^\prime$ and $3$ intersect, (\textit{ii})  $\psi=0.6978\pi$ where branches $2$ and $3$ intersect, (\textit{iii}) $\psi=1.143\pi$ where branches  $1$ and $3$ intersect and (\textit{iv})  $\psi=1.6978\pi$ where branches  $2^\prime$ and $3$ intersect. For $18$ values
of $\psi\in[0,2\pi)$,   a pair of SPP waves can be simultaneously excited  with the same attenuation rates in the interface plane but with different phase speeds.

SPP-wave propagation on branches  $1$, $1^\prime$, $2$, $2^\prime$, $3$, $4$ for $0.311\psi\leq\psi\leq 0.482\pi$, and $4^\prime$ for $1.311\pi\leq\psi\leq 1.482\pi$ in Fig.~\ref{fig:reqT205}
occurs with $\vph>\co$. Phase  speeds lower than the speed of light in free space are associated with SPP waves on
branch $4$ for $0.482\pi\leq \psi\leq 0.739\pi$,   branch $4^\prime$ for $1.482\pi\leq \psi\leq 1.739\pi$, $5$, 
and $5^\prime$. As revealed in  Fig.~\ref{fig:propdT205}, $\propdist > 2$~mm for every branch except branch $3$. Moreover, as $\psi$ increases, branch  $3$ at 205~K shows nearly constant values of $\propdist$.

Data on AEDs, relative phase speeds,  and propagation distances at $T=205$~K are provided in \tref{tab:T205}. The longest propagation distances are at least one order magnitude less than they are for $T=180$ and $200$~K. In fact, 
the longest propagation  distance at $T=205$~K is 416.64~mm, which arises on  branches $4$ and $4^\prime$.

	%%%%%%%%%%% BEGIN TABLE 3 %%%%%%%%%%%%%%%%%%%%%%%%%%%%%%%%%%
	\begin{table*}[h]
		\caption{Angular existence domain, minimum and maximum values of normalized phase speed $\vph/\co$, and minimum and maximum values  of propagation distance $\propdist$ for SPP waves guided by the InSb/SCM interface when $T=205$~K. }
		\centering
		\begin{tabular}{c|cccd{2}d{2}}
			\hline\hline
			Branch & AED ($\psi/\pi$) & $v_{ph}^{\tond{min}}/c_o$ & $v_{ph}^{\tond{max}}/c_o$ &
			 \multicolumn{1}{c}{$\propdist^{\tond{min}}$~(mm)} & \multicolumn{1}{c}{$\propdist^{\tond{max}}$~(mm)}\\ [0.5ex]
			\hline
			$ 1 $ & $\quadr{0.622,1.383}$     & $1.0639 $ & $1.2922$ & $3.23$  &$293.64$ \\ \hline
			$ 1^\prime $ & $\quadr{0,0.383}\cup\quadr{1.622,2}$     & $1.0639 $ & $1.2922$ & $3.23$  &$293.64$ \\ \hline
			$ 2 $ & $\quadr{0.317,0.928}$ & $1.0683 $ & $1.1589$ & $5.60$  & $318.31$ \\ \hline
			$ 2^\prime $ & $\quadr{1.317,1.928}$ & $1.0683 $ & $1.1589$ & $5.60$  & $318.31$ \\ \hline
			$ 3 $ & $\quadr{0,2}$         & $1.0559 $ & $1.1574$ & $0.12$  & $0.13$ \\ \hline
			$ 4 $ & $\quadr{0.311,0.739}$ & $0.9672 $ & $1.0338$ & $5.28$  &$416.64$ \\ \hline
			$ 4^\prime $ & $\quadr{0.311,0.739}$ & $0.9672 $ & $1.0338$ & $5.28$  &$416.64$ \\ \hline
			$ 5 $ & $\quadr{0.361,1.061}$     & $0.7152 $ & $0.8162$ & $2.31$  &$341.53$ \\ \hline
			$ 5^\prime $ & $\quadr{0,0.061}\cup\quadr{1.361,2}$     & $0.7152 $ & $0.8162$ & $2.31$ &$341.53$ \\ \hline
		\end{tabular}
		\label{tab:T205}
	\end{table*}
	%%%%%%%%%%% END TABLE 3 %%%%%%%%%%%%%%%%%%%%%%%%%%%%%%%%%%%%

%%%%%%%%%%%%%%%%% Figure 8 begins %%%%%%%%%%%%%%%%
\begin{figure}[h!]
	\centering
	\includegraphics[width=0.8\linewidth]{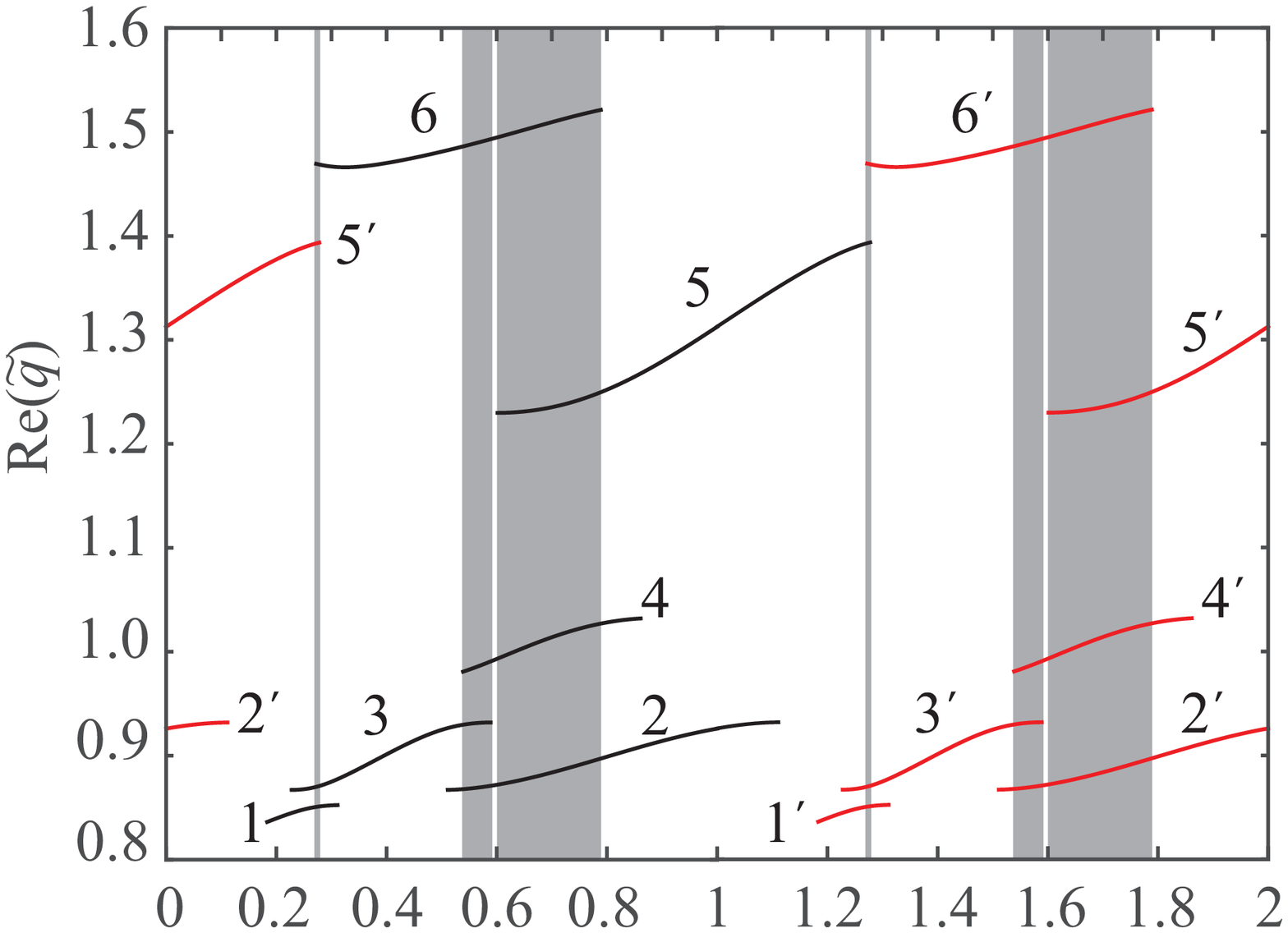}
	\caption{As \fref{fig:reqT180} except that  $T=220$~K ($\eps_{th} =  -13.66 + 2.44
		i$).}
	\label{fig:reqT220}
\end{figure}
%%%%%%%%%%%%%%%%% Figure 8 ends %%%%%%%%%%%%%%%%

%%%%%%%%%%%%%%%%% Figure 9 begins %%%%%%%%%%%%%%%%
\begin{figure}[h!]
	\centering
	\includegraphics[width=0.8\linewidth]{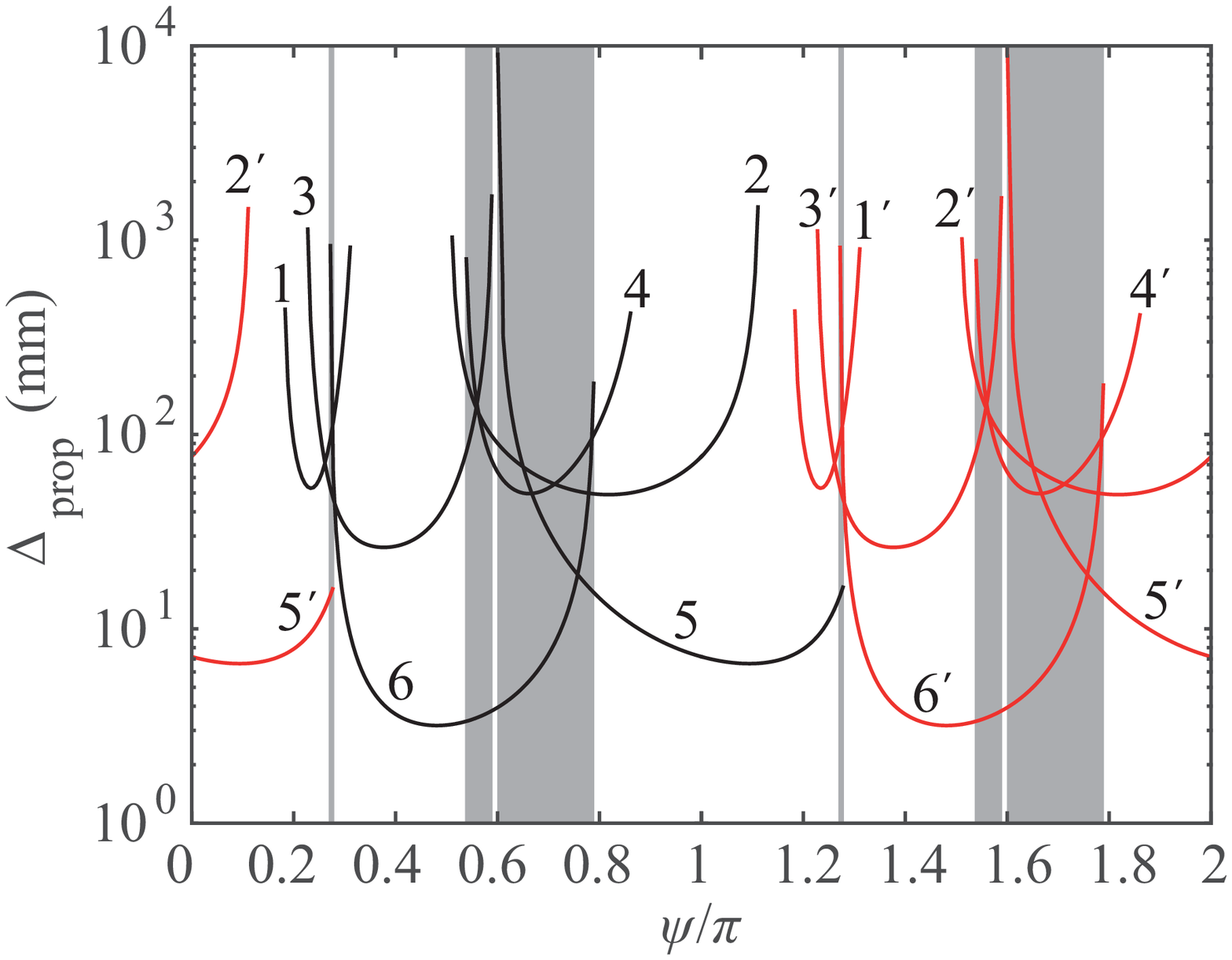}
	\caption{ As \fref{fig:propdT180} except that  $T=220$~K ($\eps_{th} =  -13.66 + 2.44
		i  $).}
	\label{fig:propdT220}
\end{figure}
%%%%%%%%%%%%%%%%% Figure 9 ends %%%%%%%%%%%%%%%%

Lastly, ${\rm Re}(\tq)$ and $\propdist$  are plotted  in  Figs.~\ref{fig:reqT220} and \ref{fig:propdT220}   as  functions of  $\psi $ for $T=220$~K.  If $q$ is a solution for a particular value of $\psi$, it is also a solution for $\psi\pm\pi$.
A comparison with Figs.~\ref{fig:reqT205} and \ref{fig:propdT205} reveals that an increase in the temperature from
$205$~K to $220$~K results in the  number of solution branches increasing to $12$. These branches are labeled $1$ to $6$ and $1^\prime$ to $6^\prime$. None of these  branches spans the entire  range of  $\psi$.

In spite of the increase in the number of solution branches at $T=220$~K, the maximum multiplicity of SPP waves is four, which arises for six angular ranges: $0.272\pi\leq \psi\leq 0.278\pi$, $0.539\pi\leq \psi\leq 0.589\pi$, $0.6\pi\leq \psi\leq 0.789\pi$, $1.272\pi\leq \psi\leq 1.278\pi$, $1.539\pi\leq \psi\leq 1.589\pi$, and $1.6\pi\leq \psi\leq 1.789\pi$. As $\psi$ increases, ${\rm Re}(\tq)$ increases for all branches. SPP waves on
branches $1$, $1^\prime$, $2$, $2^\prime$, $3$, $3^\prime$, $4$ for $0.539\leq \psi\leq 0.632\pi$, and on branch $4^\prime$ for $1.539\leq \psi\leq 1.632\pi$ in Fig.~\ref{fig:reqT220}  have  phase speeds greater than the speed of light in vacuum. 

No  intersections of branches can be seen in \fref{fig:reqT220}. Therefore, it is not possible to excite two SPP waves with the same phase speed for propagation in any specific direction in the interface plane, when $T=220$~K. For $22$ values
of $\psi\in[0,2\pi)$,   a pair of SPP waves can be simultaneously excited  with the same attenuation rates in the interface plane but, of course, with different phase speeds.

Data on AEDs, relative phase speeds,  and propagation distances at $220$~K are provided in \tref{tab:T220}. 
Long-range propagation of SPP waves is possible at 220~K, with the highest value $8.74$~m of $\propdist$ arising on branches $5$ and $5^\prime$.

%%%%%%%%%%% BEGIN TABLE 4 %%%%%%%%%%%%%%%%%%%%%%%%%%%%%%%%%%
\begin{table*}[h]
	\caption{Angular existence domain , minimum and maximum values of normalized phase speed $\vph/\co$,  and minimum and maximum values of propagation distance $\propdist$ for  SPP waves guided by the InSb/SCM interface when $T=220$~K.}
	\centering
	\begin{tabular}{c|cccd{2}d{2}}
		\hline\hline
		 Branch  & AED ($\psi/\pi$) & $v_{ph}^{\tond{min}}/c_o$ & $v_{ph}^{\tond{max}}/c_o$ &
		 \multicolumn{1}{c}{$\propdist^{\tond{min}}$~(mm)} & \multicolumn{1}{c}{$\propdist^{\tond{max}}$~(mm)}\\ [0.5ex]
		\hline
		$ 1$ & $\quadr{0.183,0.311}$ & $1.1731 $ & $1.1959 $ & $52.91 $ &$918.91$ \\ \hline
		$ 1^\prime$ & $\quadr{1.183,1.311}$ & $1.1731 $ & $1.1959 $ & $52.91 $ &$918.91$ \\ \hline
		$ 2$ & $\quadr{0.511,1.111}$     & $1.0730 $ & $1.1533 $ & $49.04 $ &$1480.62$ \\ \hline
		$ 2^\prime$ & $\quadr{0,0.111}\cup\quadr{1.511,2}$     & $1.0730 $ & $1.1533 $ & $49.04 $ &$1480.62$ \\ \hline
		$ 3$ & $\quadr{0.228,0.589}$ & $1.0730 $ & $1.1534 $ & $26.25 $ &$1684.18$ \\ \hline
		$ 3^\prime$ & $\quadr{1.228,1.589}$ & $1.0730 $ & $1.1534 $ & $26.25 $ &$1684.18$ \\ \hline
		$ 4$ & $\quadr{0.539,0.861}$ & $0.9690 $ & $1.0195 $ & $49.61 $ &$799.77$ \\ \hline
		$ 4^\prime$ & $\quadr{1.539,1.861}$ & $0.9690 $ & $1.0195 $ & $49.61 $ &$799.77$ \\ \hline
		$ 5$ & $\quadr{0.6,1.278}$       & $0.7176 $ & $0.8132 $ &  $6.62 $ &$8744.78$ \\ \hline
		$ 5^\prime$ & $\quadr{0,0.278}\cup\quadr{1.6,2}$     & $0.7176 $ & $0.8132 $ &  $6.62 $ &$8744.78$ \\ \hline
		$ 6$ & $\quadr{0.272,0.789}$ & $0.6573 $ & $0.6821 $ &  $3.18 $ &$936.21$ \\ \hline
		$ 6^\prime$ & $\quadr{1.272,1.789}$ & $0.6573 $ & $0.6821 $ &  $3.18 $ &$936.21$ \\ \hline
	\end{tabular}
	\label{tab:T220}
\end{table*}
%%%%%%%%%%% END TABLE 4 %%%%%%%%%%%%%%%%%%%%%%%%%%%%%%%%%%%%

\subsection{Power profiles}

%%%%%%%%%%%%% Figure 10 begins %%%%%%%%%%%%%
\begin{figure}[h]
	\centering
	\includegraphics[width=0.9\linewidth]{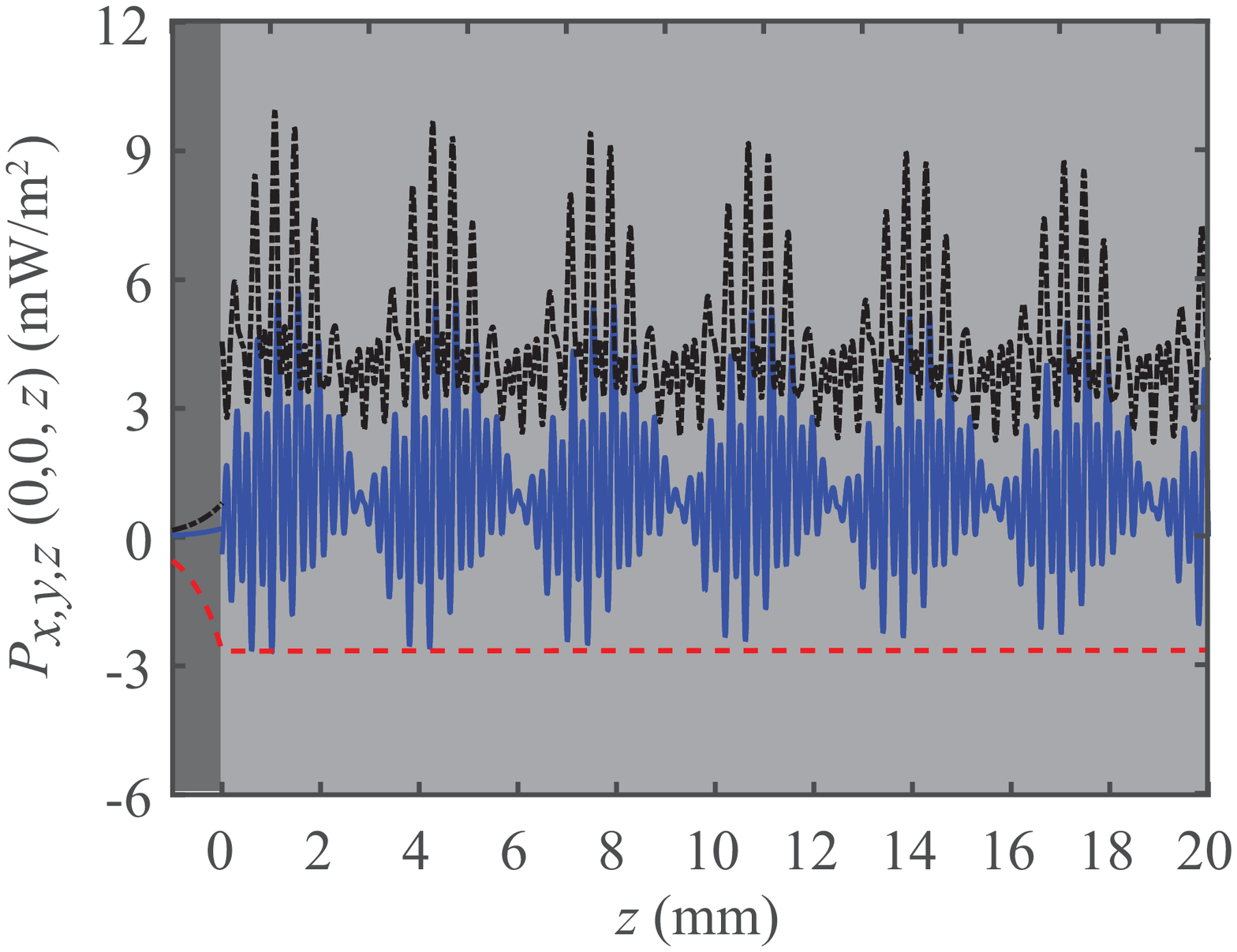}
	\caption{Cartesian components  $P_x(0,0,z)$ (blue solid lines), $P_y(0,0,z)$ (black dashed-dotted lines), and $P_z(0,0,z)$ (red dashed lines) of $\#P(0,0,z)$ plotted  against $z$ for the DT surface wave on branch~$2^\prime$ in  
	Figs.~\ref{fig:reqT180}  and \ref{fig:propdT180}. Calculations were made with $\psi=0.422\pi$
	and  $T=180$~K. The normalization protocol for these graphs is explained in a predecessor paper~\cite{ChJOPT}}
	\label{fig:HighDelta}
\end{figure}
%%%%%%%%%%%%% Figure 10 ends %%%%%%%%%%%%%

In order to further illuminate the characteristics of the ESWs, we plotted the Cartesian components of $\#P(0,0,z)$  along the $z$ axis. As an illustrative example, \fref{fig:HighDelta} presents the  spatial profile of  $\#P(0,0,z)$  
for the  DT surface wave on  branch~$2^\prime$ for $\psi=0.422\pi$ when $T=180$~K.  This ESW has the longest propagation distance ($9.95$~m) for all investigations reported in this paper.
The energy of the DT surface wave exists almost entirely in the half space occupied by the SCM; i.e., the magnitude of $\#P(0,0,z)$ is very small for  $z \in (-1, 0]$~mm, and vanishingly small for  $z < -1$~mm.
The spatial periodicity of the SCM is reflected in the plots of the $x$ and $y$ components of  $\#P(0,0,z)$ for $z>0$. Also, the DT surface wave in the SCM decays so slowly that 50 SCM periods are needed to bring about an appreciable  decay of $\#P\tond{0,0,z}$ in the half space $z>0$.

\begin{figure}[h]
	\centering
	\includegraphics[width=0.8\linewidth]{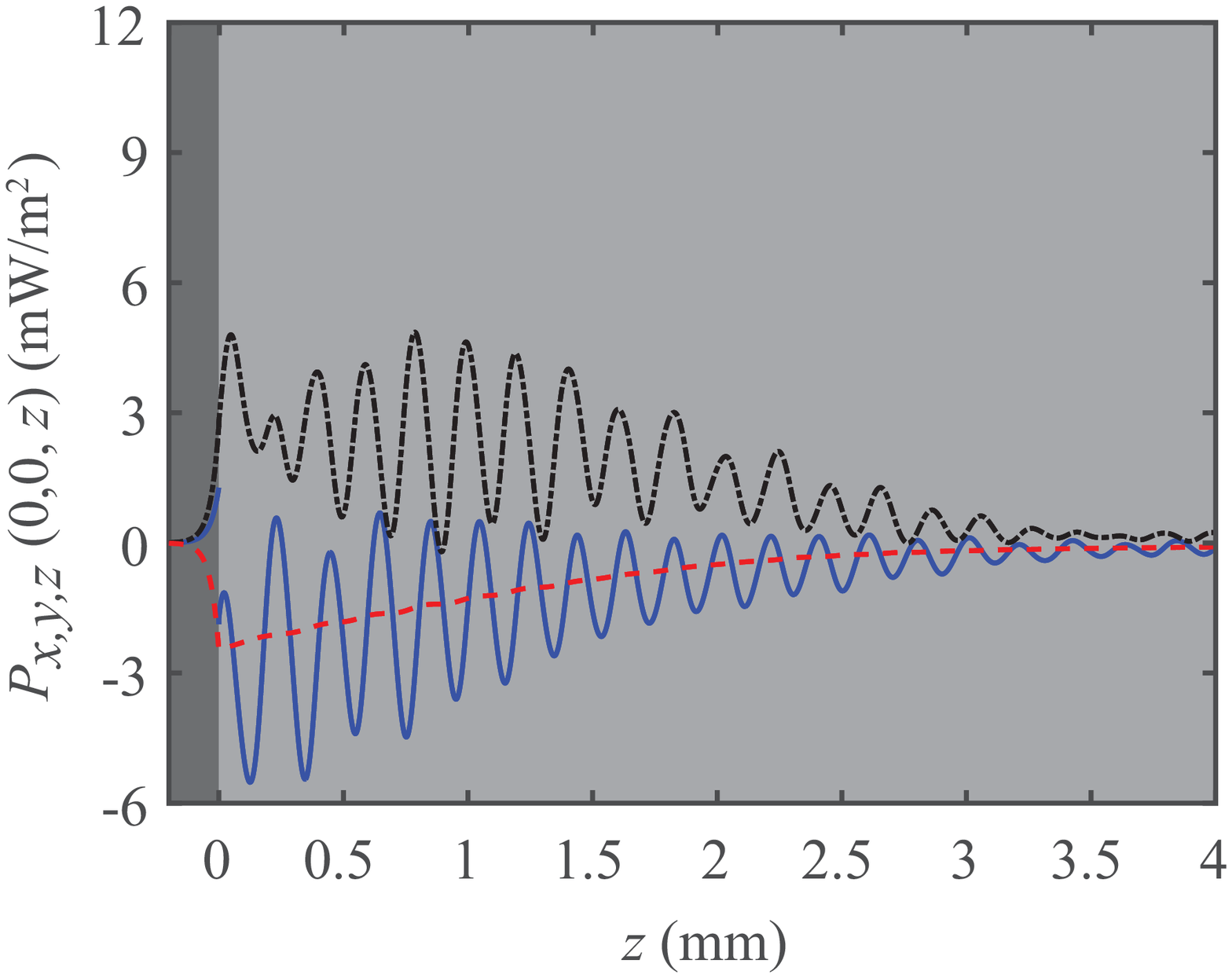}
		\caption{As Fig.~\ref{fig:HighDelta} except that 
		the Cartesian components of  $\#P(0,0,z)$ are plotted against $z$ for the SPP wave on branch 2
		in Figs.~\ref{fig:reqT205} and \ref{fig:propdT205},
		when $T=205$~K  
	and $\psi=0.6978\pi$. }
	\label{fig:cross2}
\end{figure}

\begin{figure}[h]
	\centering
	\includegraphics[width=0.8\linewidth]{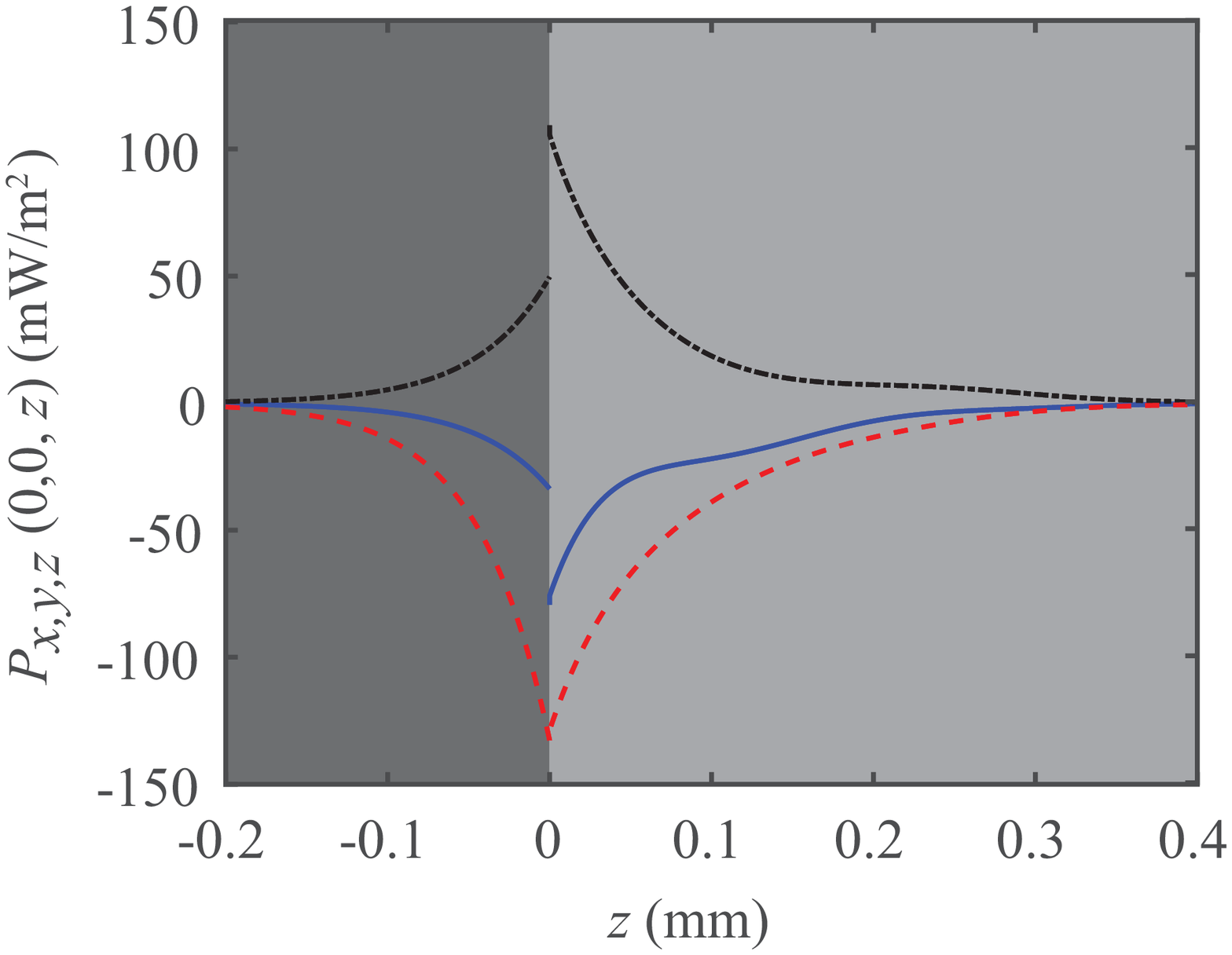}
			\caption{
			As Fig.~\ref{fig:cross2} except that		the Cartesian components of  $\#P(0,0,z)$ are plotted against $z$
			for the SPP wave on branch 3.}
	\label{fig:cross3}
\end{figure}

Figures~\ref{fig:cross2} and \ref{fig:cross3}  present the  spatial
profile of  $\#P(0,0,z)$  
for the  SPP waves on  branches~$2$ and $3$, respectively, for $\psi=0.6978\pi$ when $T=205$~K. 
At this value of
$ \psi $,  branches  $2$ and $3$ intersect in Fig.~\ref{fig:reqT205}.
The differences in their spatial profiles is striking even though both SPP waves have the same phase speed.

In Fig.~\ref{fig:cross2}, the SPP wave exists almost entirely in the  half space occupied by the SCM; i.e., the magnitude of $\#P(0,0,z)$ is negligible for $z<0$ relative to that for $z>0$.
Approximately 10 SCM periods are needed in order for the fields  to decay appreciably in the half space $z>0$.

%%%%%%%%%%%%% Figure 11 begins %%%%%%%%%%%%%
%%%%%%%%%%%%% Figure 11 ends %%%%%%%%%%%%%

%%%%%%%%%%%%% Figure 12 begins %%%%%%%%%%%%%
%%%%%%%%%%%%% Figure 12 ends %%%%%%%%%%%%%

In contrast,  the energy  of the SPP wave
is distributed across both the InSb material and the SCM   in \fref{fig:cross3}.  The components of 
$\#P(0,0,z)$ in \fref{fig:cross3} decay more rapidly in the SCM than they do in \fref{fig:cross2}. Indeed, the amplitude of  $\#P\tond{0,0,z}$ in \fref{fig:cross3} decays appreciably in only one SCM period in the $z>0 $ half space. As the energy  of the SPP wave is squeezed into much less space  in 
\fref{fig:cross3} as compared to \fref{fig:cross2},
 the maximum amplitudes of  $\#P\tond{0,0,z}$ are much greater  (about 100  $ \mbox{mW/m}^2$)  in 
\fref{fig:cross3} as compared to \fref{fig:cross2}.
Also, the plots of the $x$ and $y$ components of  $\#P(0,0,z)$ for $z>0$ in  \fref{fig:cross3} 
do not display high-frequency fluctuations, which
constrasts starkly
with the corresponding plots in Figs.~\ref{fig:HighDelta} and \ref{fig:cross2}.

\begin{figure*}[h]
	\centering
	\includegraphics[width=0.9\linewidth]{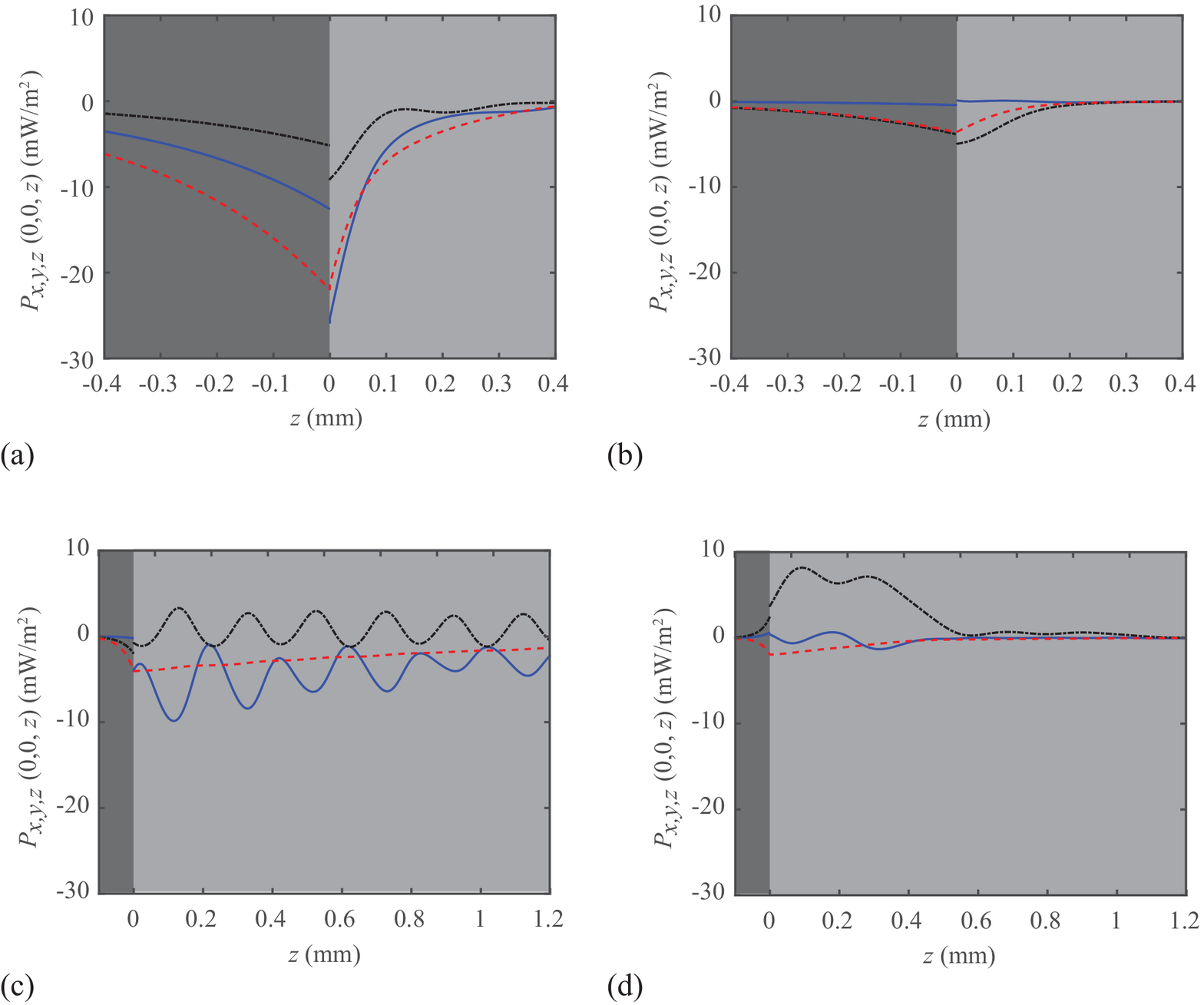}
			\caption{\blue{As in Fig.~\ref{fig:HighDelta}, Cartesian components  of $\#P(0,0,z)$ plotted  against $z$ for $T=200$~K  in (a) and (b) and $205$~K  in (c) and (d). Further details are provided in the text. }}
	\label{fig13}
\end{figure*}

\blue{
From Figs.~\ref{fig:cross2} and \ref{fig:cross3},
it may be inferred  that for SPP waves with phase speed greater than $\co$, some SPP waves are much more tightly bound to the interface than others. To explore this issue further, we turn to Fig.~\ref{fig13} wherein the
Cartesian components of $\#P(0,0,z)$  are plotted along the $z$ axis for the following cases involving $v_{ph} \lessgtr \co$ for both DT surface waves and SPP waves. 
Fig.~\ref{fig13}(a) corresponds to the branch-2 solution in Figs.~\ref{fig:reqT200}
and \ref{fig:propdT200}
at $ \psi=1.1444 \pi$. Since $T=200$ K, the ESW is a DT surface wave. Here $\tilde{q}=0.8301+i 0.4087$. The corresponding phase speed is $v_{ph} =
1.2047 \co$ and the DT surface wave is tightly bound to the interface.
Fig.~\ref{fig13}(b) is the same as Fig.~\ref{fig13}(a) except for $ \psi= 1.4500 \pi$. Here $\tilde{q}=1.2499+i 0.2731$. The corresponding phase speed is $v_{ph} =
0.8000 \co$ and the DT surface wave is tightly bound to the interface.
Fig.~\ref{fig13}(c) corresponds to the branch 1 solution in Figs.~\ref{fig:reqT205}
and \ref{fig:propdT205}
at $ \psi= 0.9722 \pi$. Since $T=205$ K, the ESW is a SPP wave. Here $\tilde{q}=0.8333+i 0.0208$. The corresponding phase speed is $v_{ph} =
1.2000 \co$ and the SPP  wave is loosely bound to the interface.
Fig.~\ref{fig13}(d) corresponds to the branch 5 solution in Figs.~\ref{fig:reqT205}
and \ref{fig:propdT205}
at $ \psi= 0.5278 \pi$. Since $T=205$ K, the ESW is a SPP wave. Here $\tilde{q}=1.24982+i 0.0236$. The corresponding phase speed is $v_{ph} =
0.8000 \co$ and the SPP  wave is somewhat loosely bound to the interface. 
}

\blue{Judging by Fig.~\ref{fig13}, as well as by  Figs.~\ref{fig:cross2} and \ref{fig:cross3},  there is no obvious relationship between the extent to which a DT surface wave or a SPP wave is bound to the interface and its phase speed relative to $\co$.
However,  a relationship is apparent
between $\Delta_{prop}$ and 
 the extent to which an ESW is bound to the interface.
 To pursue this further, let us
 introduce  $e$--folding distances in the InSb half-space and in the SCM half-space, namely $\Delta_{InSb}$ and $\Delta_{SCM}$ respectively,
 to
  characterize the 
decay of the electric field amplitude of  ESWs in the $\pm z$ directions. That is, at a distance $\Delta_{InSb}$ from the interface the magnitude of the electric field has decayed by a factor of $1/e$ in the InSb half space, and $\Delta_{SCM}$ is similarly defined for the SCM half space.
  For the numerical examples considered here, 
 every SPP wave that has $\Delta_{prop} \gtrapprox 3$ mm,
regardless of its phase speed, has a relatively large value of $\Delta_{InSb}$ and of $\Delta_{SCM}$ and therefore is loosely bound to the interface.
The same relationship also holds for DT surface waves that have  $\Delta_{prop} \gtrapprox 0.5$ mm. Some representative data are presented in Table~\ref{tab5} in support of this finding.  
}

%%%%%%%%%%%%% Figure 13 begins %%%%%%%%%%%%%
%%%%%%%%%%%%% Figure 13 ends %%%%%%%%%%%%%

%%%%%%%%%%% BEGIN TABLE 5 %%%%%%%%%%%%%%%%%%%%%%%%%%%%%%%%%%
\begin{table*}[h]
	\caption{\blue{The $e$-folding distances $\Delta_{InSb}$ and $\Delta_{SCM}$, normalized phase speed $\vph/\co$,  propagation distance $\propdist$, and ratio Im$(\tq)$/Re$(\tq)$ for  a selection of DT surface waves and SPP waves. Further details are provided in the text.}} \vspace{2mm}
	\centering
	\begin{tabular}{cccccccc}
		Branch &  $T\ (K)$ & $\psi/\pi$ & $\Delta_{InSb}$~(mm)  & $\Delta_{SCM}$~(mm)& $v_{ph}/c\ped0$ & $\propdist$~(mm) & Im$(\tq)$/Re$(\tq)$\\
		\hline\hline
		3 & 180 & 0.572 & 4.564 & 0.289 & 0.79941 & 0.69204 & 0.09192\\
		1 & 180 & 0.356 & 1.740 & 1.853 & 1.20005 & 1.30562 & 0.07314\\
		$2^\prime$ & 200 & 0.150 & 0.638 & 0.257 & 1.19804 & 0.19385 & 0.49179\\
		$2^\prime$ & 200 & 0.450 & 0.490 & 0.229 & 0.80006 & 0.29139 & 0.21850\\
		1 & 205 & 0.972 & 0.069 & 2.215 & 1.20048 & 3.82584 & 0.02497\\
		5 & 205 & 0.528 & 0.056 & 0.531 & 0.80012 & 3.37193 & 0.01888\\
		5 & 220 & 0.789 & 0.020 & 0.593 & 0.80021 & 15.37967 & 0.00414\\
		1 & 220 & 0.183 & 0.021 & 2.825 & 1.19590 & 440.55512 & 0.00022\\
		\hline
	\end{tabular} 
		\label{tab5}
\end{table*}
%%%%%%%%%%% END TABLE 5 %%%%%%%%%%%%%%%%%%%%%%%%%%%%%%%%%%

\section{Concluding remarks}\label{sec:cr}

The effect of changing temperature on the propagation of  ESWs  guided by the planar interface of  InSb
 and a  SCM was numerically investigated in the terahertz frequency regime. On raising the temperature,   InSb  is transformed
 from a dissipative dielectric material to a \blue{dissipative} plasmonic material. Consequently, the ESWs transmute from DT surface  waves to SPP waves.
The dispersion relation, arising from the solution of  a canonical boundary-value problem, can yield multiple values for the wavenumber $q$ for any  propagation direction.
A multiplicity of  ESWs was found. To be specific, for the particular scenarios considered here, as many as four DT surface waves and up to five SPP  waves may be excited for certain propagation directions. The propagation distances of some ESWs  are remarkably high, reaching in a few cases lengths of some meters;
 in such instances
 the  energy of the ESW is almost entirely  within the SCM.
\blue{While the foregoing analysis clearly demonstrates that  ESWs of a wide variety of different natures  can be excited, depending upon temperature, the reasons for this diversity of natures are not revealed. For example, the reasons for the sharp contrast between the power profiles represented in Figs.~\ref{fig:cross2}
and \ref{fig:cross3} for SPP waves at the same temperature and with the same propagation direction are not clear.
Further analysis is required to shed light on this matter.}

For certain propagation directions, simultaneous excitation of two ESWs with the same phase speeds but different propagation distances is  possible.
In a similar vein, two ESWs with the same propagation distances but different phase speeds may  be 
simultaneously excited for certain propagation directions. \blue{The simultaneous excitation of such ESWs does not infer coupling between those ESWs.
}

\blue{ Clearly, the canonical boundary-value problem considered herein, based on two infinite half spaces, is  not practically implementable. However, it provides valuable insights into the essential characteristics of ESW excitation that may be 
reasonably expected in more realistic scenarios.
 For example, qualitatively similar surface-wave characteristics may be anticipated for 
 the practically implementable
grating--coupled and waveguide--coupled configurations in the case of Dyakonov--Tamm surface waves, as well as for the
grating--coupled, waveguide--coupled, and prism--coupled configurations in the case of SPP waves \cite{PMLbook}. }

\vspace{10pt}

{\bf Acknowledgement.}
A. Lakhtakia thanks the Charles Godfrey Binder Endowment at the Pennsylvania State University for ongoing support of his research.

\end{document}